\documentclass{ws-ijmpb}
\usepackage{graphicx,bm}

\newcommand{\kf}{k_{\rm F}}
\newcommand{\be}{\begin{equation}}
\newcommand{\ee}{\end{equation}}
\newcommand{\q}{{\bf q}}
\newcommand{\qp}{{\bf q}'}
\newcommand{\Pt}{\bf P}
\newcommand{\p}{\bf p}
\newcommand{\pp}{{\bf p}'}
\newcommand{\la}{\Lambda}
\newcommand{\si}{{\bm \sigma_1}}
\newcommand{\sip}{{\bm \sigma_2}}
\newcommand{\vlk}{V_{{\rm low}\,k}}

\begin{document}

\markboth{Achim Schwenk}{Nuclear interactions from the renormalization group}
\catchline{}{}{}{}{}
\title{Nuclear interactions from the renormalization group}

\author{Achim Schwenk}
\address{Department of Physics, The Ohio State University,
Columbus, OH 43210~\footnote{
Present address: Nuclear Theory Center, 
Indiana University, Bloomington, IN 47408.}}
\maketitle


\begin{abstract}
We discuss how the renormalization group can be used to derive
effective nuclear interactions. Starting from the model-independent
low-momentum interaction $\vlk$, we successively integrate out 
high-lying particle and hole states from momentum shells around 
the Fermi surface as proposed by Shankar. The renormalization
group approach allows 
for a systematic calculation of induced interactions and yields
similar contributions to the scattering amplitude as the two-body
parquet equations. We review results for the $^1$S$_0$ and $^3$P$_2$ 
superfluid pairing gaps as well as the spin dependence of effective 
interactions in neutron matter. Implications for the cooling of 
neutron stars are discussed.
\end{abstract}


\section{Introduction}

Conventional precision nucleon-nucleon (NN) interactions are
well-constrained by two-nucleon scattering data only for laboratory 
energies $E_{\rm lab} \lesssim 350 \, {\rm MeV}$. As a consequence, 
details of nuclear forces are not constrained for relative momenta 
$k > 2.0 \, {\rm fm}^{-1}$ or for relative distances $r < 0.5 \, {\rm fm}$.
All these NN potentials have repulsive cores, which lead to significant 
probing of the model-dependent high-momentum components in few and 
many-body applications. This model dependence, when not compensated 
by three-body interactions, results in the Tjon and Coester lines.

Using the renormalization group (RG), we have integrated out the 
high-momentum modes above a cutoff $\la$ in momentum space. The 
resulting low-momentum interaction, called $\vlk$, only has momentum 
components below the cutoff and evolves with $\la$ so that all 
low-energy two-body observables below the cutoff (phase shifts and 
deuteron binding energy) are invariant. We have shown that for 
$\la \lesssim 2.0 \, {\rm fm}^{-1}$, all NN potentials that 
fit the scattering data and include the same 
long-distance pion physics lead to the same ``universal'' low-momentum 
interaction $\vlk$.\cite{Vlowk1,Vlowk2} When $\vlk$ is augmented by
a low-momentum three-nucleon (3N) force, which regulates $A=3,4$ binding 
energies, we find that the 3N parts are perturbative for
$\la \lesssim 2.0 \, {\rm fm}^{-1}$.\cite{Vlowk3NF} By perturbative
we mean $\langle \Psi^{(3)} | V_{\rm 3N} | \Psi^{(3)} \rangle \approx
\langle \Psi^{(2)} | V_{\rm 3N} | \Psi^{(2)} \rangle$, where $| \Psi^{(n)}
\rangle$ are exact solutions including up to $n$-body forces.

\begin{figure}[t]
\begin{center}
\includegraphics[scale=0.4,clip=]{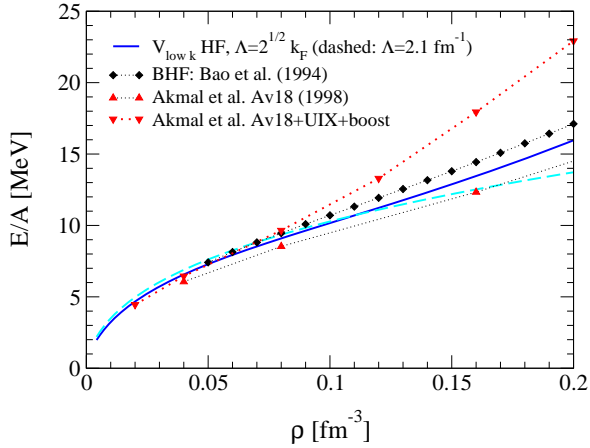}
\hspace*{2mm}
\begin{minipage}{4.5cm}
\vspace*{-7cm}
\caption{\label{neutmat}Comparison of the $\vlk$ HF EoS 
for pure neutron matter with Brueckner-Hartree-Fock (BHF) results 
using the Bonn potential and results of Akmal {\it et al.} obtained
using chain summation methods, for references see.$^4$ Calculations
indicate that nuclear matter may be perturbative with low-momentum 
interactions.$^5$ Contrary to all other microscopic interactions, 
$\vlk$ yields bound nuclei on the HF level.$^6$}
\end{minipage}
\end{center}
\vspace*{-5mm}
\end{figure}

Since $\vlk$ does not have a strong core at short distances, it can
be used directly in many-body applications without a $G$ matrix
resummation. Fig.~\ref{neutmat} shows
the equation of state (EoS) of pure neutron matter obtained in the 
Hartree-Fock (HF) approximation.\cite{RGnm} In symmetric nuclear matter, 
saturation is due to the 3N force.\cite{nuclmat}

\section{Renormalization group approach to neutron matter}

A central challenge in nuclear many-body theory lies in understanding
nuclear data and predicting input for astrophysical simulations from
microscopic nuclear interactions. In these proceedings, we review
results for the superfluid pairing gap, where many-body correlations 
are crucial even in the perturbative limit, due to the singular 
dependence of the gap on the interaction. For dilute neutrons
interacting through an attractive S-wave scattering length $a_{\rm S} 
< 0$ only, the pairing gap is given by\cite{Gorkov}
\begin{eqnarray}
\Delta &=& \frac{8}{e^2} \, \varepsilon_{\rm F} \exp \biggr\{
\, {\rm const.} \, \biggr( \,
\begin{minipage}{4.4cm}
\includegraphics[scale=0.5,clip=]{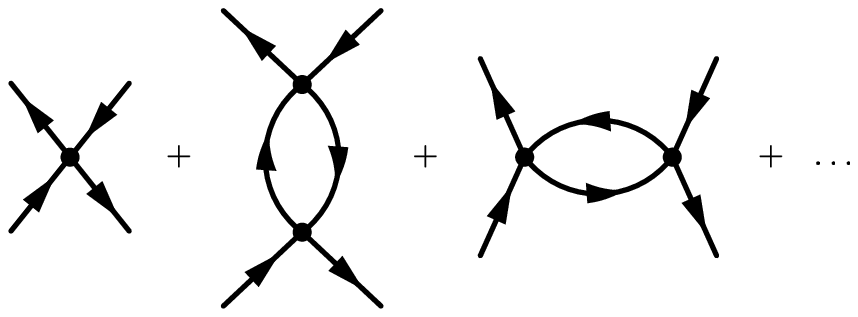}
\end{minipage}\biggr)^{-1} \biggr\} \nonumber \\
&=& \frac{8}{e^2} \, \varepsilon_{\rm F} 
\exp\biggl(\frac{2}{\pi \kf a_{\rm S}} + \log(0.45) + {\mathcal O}(\kf 
a_{\rm S})\biggr) ,
\end{eqnarray}
where the second-order diagrams are projected on S-wave and $e$ is the
exponential constant. Thus, particle-hole (ph) polarization effects 
reduce the pairing gap by a factor $(4e)^{-1/3} \approx 0.45$ compared 
to the mean-field BCS estimate, due to long-range spin fluctuations. 
In neutron matter, it is therefore crucial to include ph correlations 
through the induced interaction for a realistic assessment of 
polarization effects.

\begin{figure}[b]
\vspace*{-3mm}
\begin{center}
\includegraphics[scale=0.4,clip=]{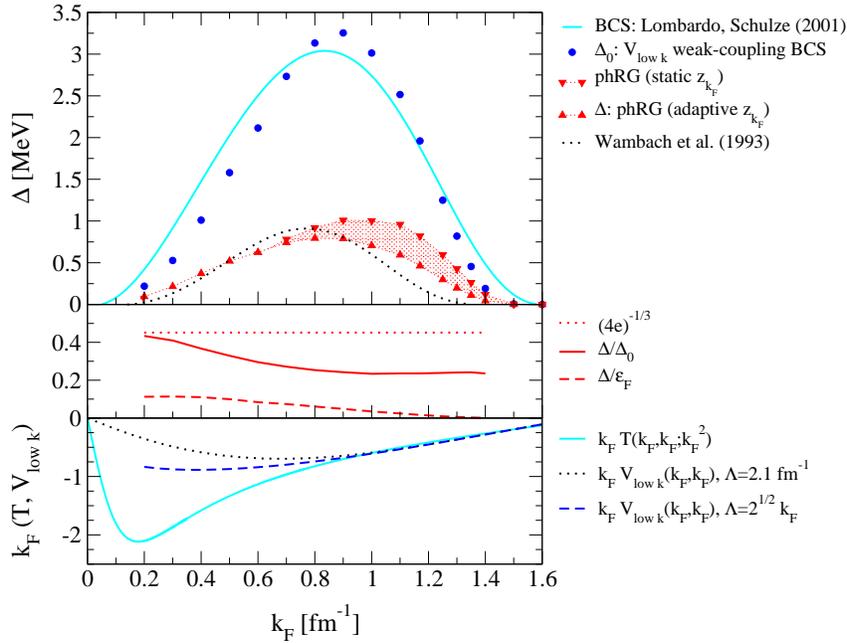}
\end{center}
\vspace*{-3mm}
\caption{\label{1s0pairing}Upper panel: Comparison of mean-field BCS
results for the $^1$S$_0$ superfluid gap to results including polarization
effects through the phRG, for references and further details, 
including the $z_{\kf}$ factor, see.$^4$ Middle
panel: Comparison of the full superfluid gap to the mean-field BCS gap
and the Fermi energy. We also show the universal extreme low-density
limit, $\Delta/\Delta_0 = (4e)^{-1/3} \approx 0.45$. Lower panel:
Comparison of dimensionless lowest-order pairing interactions.}
\end{figure}

In many-body systems, the RG provides a systematic tool to construct
non-perturbative effective interactions among valence nucleons. We 
have applied the RG to neutron matter, restricting the effective 
interaction to low-lying states in the vicinity of the Fermi 
surface.\cite{RGnm} This follows the RG approach to interacting Fermi 
systems proposed by Shankar.\cite{Shankar} The method is widely used 
in condensed matter physics to study the interference of different
instabilities, especially in the context of the 2d Hubbard 
model.\cite{2dHubbard} Starting from $\vlk$ in the full space, the RG 
generates the induced interaction, i.e., screening and vertex corrections, 
which contribute to the quasiparticle interaction and the low-energy 
scattering amplitude in the effective theory defined for
particle/hole modes within a momentum shell of width $\la$ 
around the Fermi surface.
At one loop, the change of the effective four-point vertex $a(\q,\qp;\la)$
is given by an RG equation, which reads diagrammatically 
\be
\begin{minipage}{7.5cm}
\includegraphics[scale=0.6,clip=]{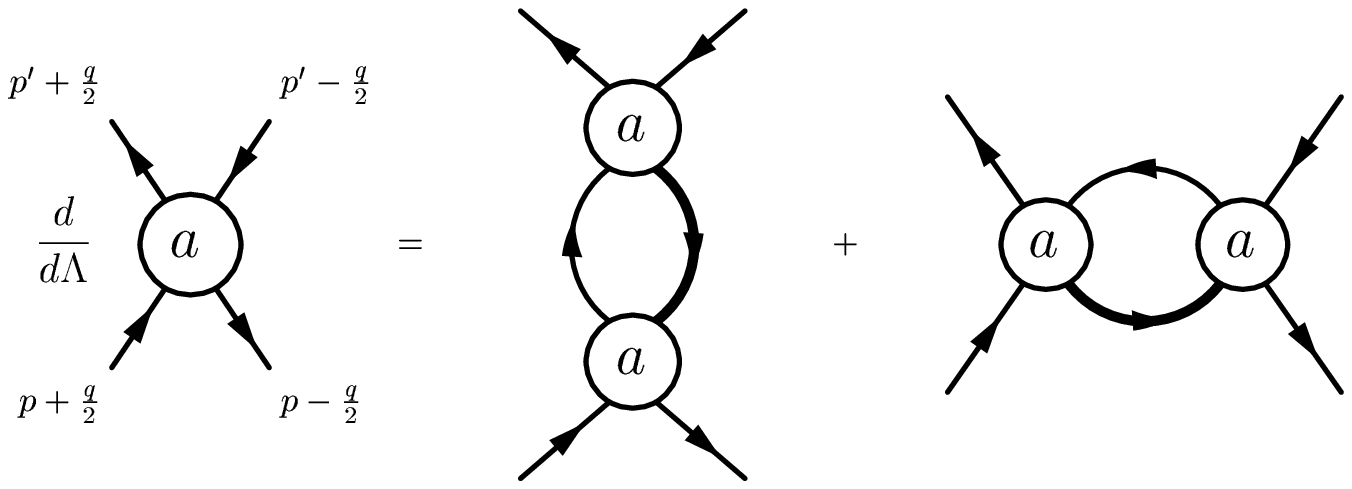}
\end{minipage}
\label{RGE}
\ee
Here, the thin lines denote intermediate states from thin momentum shells, 
which are integrated out successively, and the thick lines denotes ``fast''
particles/holes with $|p_i-\kf| \geqslant \la$. The RG treats the dependence
on the momenta $\q$ and $\qp=\p-\pp$ on an equal footing and maintains
the symmetries of the scattering amplitude. Currently, we treat pairing
correlations explicitly in weak-coupling BCS theory, 
after the RG is evolved to
the Fermi surface. Future directions will include the BCS channel in
the RG equation. On the Fermi surface, $\q$, $\qp$ and $\Pt=\p+\pp$
are orthogonal, and therefore, we approximate $a(\q,\qp;\la) = 
a(q^2,q'^2;\la)$ in the RG to extrapolate off the Fermi surface.
On the $\vlk$ level, the $\q \cdot \qp$ dependence is small.

The efficacy of the RG lies in including many-body correlations from
successive momentum shells, on top of an effective interaction with 
particle/hole polarization effects from all previous shells. By solving 
Eq.~(\ref{RGE}) iteratively, it can be seen that the RG builds up 
correlations similar to the two-body parquet equations, see 
also.\cite{indint} 

\begin{figure}[t]
\vspace*{-1mm}
\begin{center}
\includegraphics[scale=0.375,clip=]{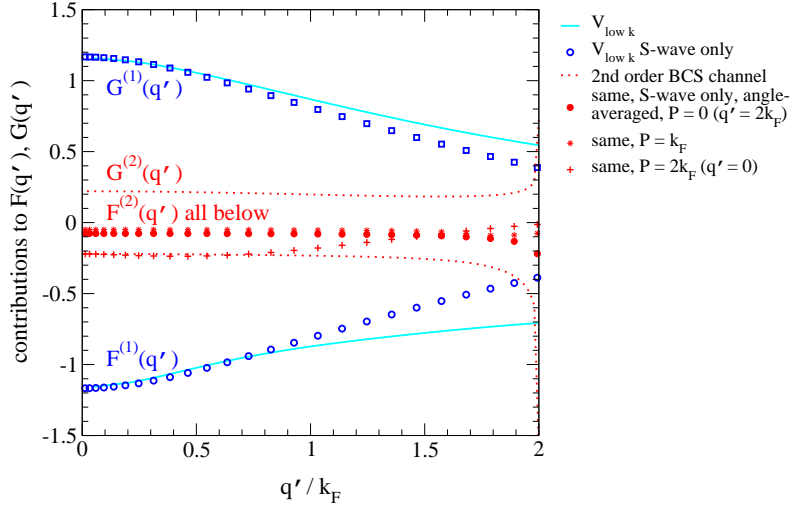}
\end{center}
\vspace*{-4mm}
\caption{\label{BCSchannel}Comparison of $\vlk$ {\scriptsize $(1)$} and 
second-order BCS channel contributions {\scriptsize $(2)$} to the scalar, 
$F(q')$, and spin-spin dependent Landau functions, $G(q')$, in 
neutron matter for $\kf=1.0 \, {\rm fm}^{-1}$. The lines are 
from$^{13}$ and the points are S-wave approximations thereof. 
The differences on the $\vlk$ level are due to contributions from 
higher partial waves. Small $q' \to 0$, where 
higher partial waves are negligible in $F^{(1)}(q')$, correspond
to $P \to 2 \kf$. Differences for $F^{(2)}(q')$ for larger $q'$ 
are therefore due to higher partial waves, angle-averaging, as well as 
hole-hole contribions, which are included in the dotted lines, but 
not the points, and are also absent for $P \to 2 \kf$. In contrast 
to free-space scattering, where the second Born term is comparable 
to $\vlk$, we find that the second-order particle-particle
contributions are small, except for low-lying pairing correlations 
($\sim \log(P/\kf)$).}
\vspace*{-3mm}
\end{figure}

In neutron matter, non-central and 3N forces are weaker, and we have 
solved the RG equations with the initial condition $a(\la=\kf) = \vlk$ 
including only scalar and spin-spin interactions, which dominate at 
low densities. In addition to the Fermi liquid parameters 
(see Fig.~6 in\cite{RGnm}), 
the RG solution yields the scattering amplitude for finite 
scattering angles, which is needed, e.g., for transport and pairing. In 
Fig.~\ref{1s0pairing}, we present our results for the $^1$S$_0$ 
pairing gap. We find a suppression of the S-wave gap due to spin 
fluctuations from $\Delta_0 \approx 3.3 \, {\rm MeV}$ to $\Delta
\approx 0.8 \, {\rm MeV}$ at maximum. Our results are similar to 
those of Wambach {\it et al.},\cite{Wambach} and our mean-field 
($\vlk$) weak-coupling gap agrees well with the BCS result. This 
is consistent with the agreement of the $\vlk$ HF EoS 
with the BHF results and shows that $\vlk$ does 
not lead to strong short-range correlations. In 
Fig.~\ref{BCSchannel}, we also compare the S-wave second-order 
particle-particle contributions to the full second-order
result.

\section{Spin-dependence of effective interactions}

\begin{figure}[t]
\vspace*{-1mm}
\begin{center}
\includegraphics[scale=0.375,clip=]{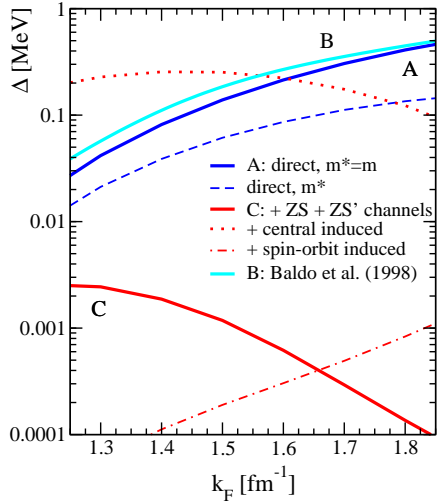}
\hspace*{2mm}
\begin{minipage}{6.0cm}
\vspace*{-7.4cm}
\caption{\label{3p2gaps}The angle-averaged $^3$P$_2$ superfluid gap
versus Fermi momentum in neutron
matter. The direct ($\vlk$) weak-coupling result for a free spectrum 
and with effective mass is compared to the gap which includes second-order 
ph polarization effects on the pairing interaction. We also 
show the modification of the gap, if only induced central or only induced 
spin-orbit effects are taken into account. As a reference, we give the 
results of Baldo {\it et al.}, which are obtained by solving the BCS 
gap equation in the coupled $^3$P$_2$--$^3$F$_2$ channel for different 
free-space interactions. As predicted by Pethick and Ravenhall, and 
confirmed by our results, superfluidity is enhanced, if only central 
induced interactions were included. For details and references see.$^{13}$}
\end{minipage}
\end{center}
\vspace*{-8mm}
\end{figure}

Non-central forces in nuclear matter lead to novel spin-dependencies
due to the presence of the Fermi sea. In addition to the standard 
forces, novel spin non-conserving effective interactions, $i (\si
- \sip) \cdot \q \times \Pt$ and $(\si \times \sip) \cdot (\qp \times
\Pt)$, are induced by screening in the ph channels and a 
novel long-wavelength tensor force $S_{12}(\Pt)$ is generated.\cite{tensor} 
Moreover, due to the coupling of the tensor and spin-orbit force to 
the strong spin-spin interaction $G_0$, we find that the tensor component
of the quasiparticle interaction and the P-wave spin-orbit pairing
force are significantly reduced in neutron matter, 
see Figs.~1 and~2 in.\cite{tensor} As illustrated
in Fig.~\ref{3p2gaps}, the screening of the spin-orbit interaction 
leads to a strong suppression of the $^3$P$_2$ superfluid gap for 
neutrons in the interior of neutron stars. Note that in vacuum the 
spin-orbit force is crucial for a realistic description of the P-wave
phase shifts and that the suppression of the gap is due to a reduction
of the pairing interaction by only $< 50\%$.

Typical interior temperatures of isolated neutron star sources are on
the order of $T \sim 10^8 \, {\rm K} \approx 10 \, {\rm keV}$ and therefore
it is possible to constrain the P-wave gaps phenomenologically through 
neutron star cooling simulations. It has been found that with free-space 
gaps $\Delta_0 \sim 0.1 \, {\rm MeV}$, neutron stars cool too
rapidly and that a consistency with the data requires $^3$P$_2$ gaps
$\Delta \lesssim 30 \, {\rm keV}$~\cite{cooling1} or 
$\Delta \lesssim 10 \, {\rm keV}$.\cite{cooling2} 
A less severe but still relevant dependence on the gap has been
found in.\cite{cooling3} The same microscopic scattering amplitudes
can be used to constrain neutrino emissivities.\cite{nnbrems} Finally,
polarization effects and novel spin-dependences are important for
magnetic susceptibilities and long-wavelength spin-isospin 
response, as discussed in.\cite{OP} 

Work is in progress to extend the RG approach to non-central
interactions and to asymmetric matter 
including neutron/proton pairing.\cite{asymmatter} In 2d, higher 
loops in the RG equation are geometrically suppressed by $\la/\kf$ 
for regular Fermi surfaces.\cite{Shankar} This may enable an error 
estimate of the one-loop truncation for low-momentum interactions,
which do not scatter strongly to high-lying (high $\la$) states. 
Finally, the RG approach allows for a consistent renormalization of 
currents/operators and an extension to effective valence shell-model 
interactions for heavy nuclei.

\vspace*{1mm}
\noindent
I am grateful to Scott Bogner, Gerry Brown, Bengt Friman, Dick Furnstahl,
Chuck Horowitz, Tom Kuo, Emma Olsson, Janos Polonyi and Chris Pethick for 
many useful discussions. This work was supported by the NSF under grant No. 
PHY-0098645 .

\end{document}